\documentclass[journal]{IEEEtran}

\usepackage{graphicx}
\usepackage{subfigure}
\usepackage{float}
\usepackage{amsmath}
\usepackage{breqn}
\usepackage{empheq}
\usepackage{url}
\usepackage{epstopdf}

\ifCLASSOPTIONcompsoc
  \usepackage[nocompress]{cite}
\else
  \usepackage{cite}
\fi

\usepackage{mathptmx} 
\usepackage{times} 
\usepackage{amssymb}  
\usepackage{upgreek}

\ifCLASSINFOpdf
\else
\fi

\begin{document}
%
\title{Building a navigable fine texture design space}
%
%
%
\author{Rebecca~Fenton~Friesen,~\IEEEmembership{Member,~IEEE,}
        Roberta~L.~Klatzky,
        Michael~A.~Peshkin,~\IEEEmembership{Senior Member,~IEEE}
        and~J.~Edward~Colgate,~\IEEEmembership{Fellow,~IEEE}
\thanks{First, third and fourth authors are with the Department
of Mechanical Engineering, Northwestern University, Evanston,
IL, 60208 USA. Contact: rebeccafentonfriesen.github.io/\#contact}
\thanks{R.L. Klatzky is with Carnegie Mellon University.}
\thanks{Manuscript received Month XX, 20XX; revised Month XX, 20XX.}}

%



\maketitle

\begin{abstract}
Friction modulation technology enables the creation of textural effects on flat haptic displays. However, an intuitive and manageably small design space for construction of such haptic textures remains an unfulfilled goal for user interface designers. 
In this paper, we explore perceptually relevant features of fine texture for use in texture construction and modification. 
Beginning with simple sinusoidal patterns of friction force that vary in frequency and amplitude, 
we define “irregularity” as a third building block of a texture pattern and show it to be a scalable feature distinct from the others using multidimensional scaling. 
Additionally, subjects' verbal descriptions of this 3-dimensional design space provide insight into their intuitive interpretation of the physical parameter changes.

\end{abstract}


%
\IEEEpeerreviewmaketitle

\section{Introduction}

\IEEEPARstart
{T}{ouchscreens} are an essential window through which we explore and control our environments, yet they provide little or no actual touch feedback to their users. Friction modulation technology allows us to build touchscreens that can provide a wide variety of tactile effects by inducing variable lateral forces on the fingertip, either via ultrasonic friction modulation \cite{watanabe1995method, winfield2007t, biet2007squeeze} or electroadhesion \cite{shultz2018application, vardar2017effect}. 
One effect particularly well suited to these friction-modulating haptic displays is texture simulation, as texture is already a highly salient characteristic of flat surfaces. Of particular interest are fine textures, which elicit primarily higher frequency vibrations on the skin during active touch.

A wide range of textures available for display could facilitate communication by helping users navigate a screen or locate icons and buttons. Haptic texture can also facilitate emotive communication, for example by simulating the sensation of stroking a friend's arm or petting a favorite dog's fur. However, building satisfying textures for these purposes is nontrivial: how can we provide texture designers a library of differentiable textures, as well as usable tools to enhance and modify them? An ongoing goal is to identify characteristics of variable friction fine textures that can be both mathematically and perceptually scaled. Such continuously variable features can then serve as our tools of enhancement and modification. 

\subsection{Defining a texture workspace}

Texture in any sensory modality, whether visual, auditory, or tactile, encompasses a wide range of length scales. These range from larger coarse features to very fine micro details, and not all of these scales should necessarily be designed and modified using the same parameter sets. 
Tactile texture specifically consists of both macro texture, detected as larger distinct features on the surface, and micro (or "fine") texture, which is detectable primarily through vibrations produced on the skin during active exploration. David Katz referred to this macro versus micro duality almost a century ago in his book "The World of Touch", an early and hugely significant introduction to tactile perception \cite{katz1925world}.
Later work has revealed that the somatosensory system has disparate methods of detecting the two types of texture via different types of mechanoreceptors, which are responsible for either lower or higher frequency content \cite{lederman2009haptic}.
Our ability to discriminate fine textures from each other is strongly correlated with differences in frequency content produced on the finger during scanning \cite{manfredi2014natural}.

In this work we are focusing on textures composed of finer length scale vibrations, applied using variable friction displays. Several studies have found that at length scales under 1mm, the relative phases of frequency components do not affect texture discrimination \cite{meyer2015modeling,cholewiak2009frequency}. This demonstrates that for fine texture, quite a bit of information is lost or combined by the somatosensory system, making these textures ripe for reduced representation. 
A primary inspiration for our approach to texture design and modification is the "Tactile Paintbrush" approach by Meyer et. al. \cite{meyer2016tactile}, which posited that any friction-varying pattern composed of length scales of 1 mm$^{-1}$ or finer could be entirely described by its magnitude spectrum, as well as a statistical characterization of how those magnitudes varied over time or space. This results in vastly fewer parameters to describe a given texture than an entire mapping of all friction values for the length of the texture, but remains an unwieldy amount (200+ numbers to adjust) for a texture designer. 

Subsequent work suggests that a detailed description of a fine texture's spectrum is probably still an over- parameterization: several studies of textures or vibrations composed of 2 frequency components demonstrate that they are quite perceptually similar to those composed of only one \cite{friesen2018single,hwang2017perceptual}. Here, a single perceived frequency, or pitch, might be a characteristic of a more complicated spectrum. 
What other simplified characteristics can we extract from a texture's spectrum that summarize its most salient perceptual qualities? 
We can look for relevant characteristics using 
multidimensional scaling, which provides a means to visualize how various mathematical parameters of a texture are reflected in a perceptual space.

\subsection{Perceptual dimensions of natural textures}

Multidimensional Scaling (MDS) is an analysis technique that visualizes perceived dissimilarities between stimuli as distances in a perceptual space. 
In perceptual space, we can look for clusters of similar-feeling stimuli, or how physical changes to those stimuli result in perceptual changes. 
We can also determine how many dimensions account for perceptual differences by looking at the amount of stress imposed when mapping all the dissimilarities into an n-dimensional space.
This analysis technique has been applied to multiple sensory modalities, such as taste and vision, to reveal relationships between a wide array of consumer products and psychophysical stimuli,
and has been used extensively in haptic texture perception since the 1990s. 

Many of the earlier studies looking at the perceptual space of texture queried the differences amongst real samples, such as wood, sandpaper, or different types of cloth. One of the first, by Hollins et. al. \cite{holliins1993perceptual}, found a 3 dimensional space with perceptual axes of rough/smooth, hard/soft, sticky/slippery. This dimensionality and types of axes are commonly found in many other studies as well, along with alternative axes such as moist/dry and cold/warm; see \cite{okamoto2012psychophysical} for a thorough overview of work on textural perceptual dimensions through 2013. Differences in primary axes found may be due to particular textures chosen to be included in a given set. 

In addition to finding perceptual axes, we often wish to know what physical properties underlie a given texture's position along a perceptual axis. 
As demonstrated by Bergmann Tiest and Kappers \cite{tiest2006analysis}, gradients pointing in the direction of greatest change of measured parameters of the textures, such as surface roughness or average kinetic friction, can be mapped onto the perceptual space. 
Work looking at the dimensionality of real textures continues to this day; e.g., Vardar et al \cite{vardar2019fingertip} recently demonstrated that the three dimensions originally found by Hollins are related to physical measurements of spectral features such as power, centroid, and DC friction levels. 

\subsection{Perceptual dimensions of textures for haptic displays}

With the advent of more sophisticated haptic displays, researchers have been able to extend the range of possible textures to artificial ones composed of applied forces and vibrations. 
One advantage of artificial textures is the opportunity to actually choose and precisely adjust physical parameters, beyond simply measuring their values. On the other hand, it is by no means clear how best to select parameters or synthesize these textures.
To date, designs selected for MDS analysis generally consist of a modest repertoire of 
features, such as only a few force values or frequency components.
For example, one study working with an electroadhesive screen generated a coarse texture set consisting of only two friction levels \cite{mun2019perceptual}. The authors varied the overall friction difference, as well as the line width and spacing for different tessellated patterns, and found that parameters related to intensity, such as line thickness and friction levels, dominated perception. Either pattern shape or density (i.e. frequency of line spacing), depending on overall intensity, was also significant. Similarly, a study applying pulses of force via a stylus varied the amplitude and timing of pulses, and found that amplitude of pulses mapped very strongly onto their 2D perceptual space, while frequency and different rhythm patterns were somewhat entwined along the same axis \cite{ternes2008designing}.

Haptic texture MDS studies, including the ones above, often look for groups or clusters of similar textures. 
Bernard et al. \cite{bernard2018harmonious} instead looked for continuous changes in engineering parameters reflected in MDS space. Their texture set, displayed on an ultrasonic friction modulating screen, consisted of pairs of harmonic frequency components.
Textures were best fit by two dimensions and varied by overall amplitude, fundamental frequency value, and inter-frequency spacing. 

Finally, we cannot automatically assume that artificial textures will have the same perceptual axes as natural ones.
Real world textures include many additional haptic cues, such as thermal variation and deformability, that are not achievable with friction modulation alone. 
Will the limited range of variable features on a haptic screen result in changes to the most salient perceptual axes?
With this in mind, Dariosecq et al. recently compared verbal descriptions of artificial textures to those used for real ones in earlier perceptual space studies \cite{dariosecq2020investigating}.
Their artificial textures consisted of sinusoidal and square variations in friction on an ultrasonic friction modulation screen. Analyzing the words subjects used to describe different patterns using factor analysis, 
they found that the rough/smooth axis was indeed preserved from real to friction modulated textures. This axis depended most strongly on waveform amplitude, as well as spatial frequency to a lesser degree.

\subsection{Parameter choice for texture workspace}

Previous research exploring the perceptual dimensionality of artificial textures always uses frequency of regularly spaced features as a controlled parameter, albeit in a variety of forms, i.e. pulse timing, tessellated pattern size, or period of sinusoidal and square waves. Intensity is also a reoccurring parameter, most often as the strength of pulses or overall friction range. Intensity is also related to wave form; square waves will feel more intense than sinusoidal due to the introduction of harmonics, which add to the power of the signal. Overall, frequency and intensity are clearly salient characteristics, and as features of a sinusoid (where intensity translates as amplitude) they constitute the simplest form of spectral information. Therefore, we also chose these two characteristics of a sinusoidal variation in friction as our first two parameters.

It is worth noting that frequency and amplitude in the fine texture regime are not necessarily perceptually orthogonal.
A better term for frequency perception would be pitch, the perceived frequency of a signal that is dependent both on actual frequency and amplitude. Georg von Bekesy, a Nobel Prize winning scientist best known for his early work in audition and pitch perception via the cochlea, also wrote extensively about tactile pitch perception, noting that perceived pitch is inversely related to amplitude \cite{von1959synchronism}. Studies looking at only frequency and amplitude as parameters often first construct iso-amplitude functions of frequency in order to keep the two features independent of each other, e.g. in  \cite{hwang2017perceptual}. While we have not done this in the following study, it will be interesting to observe the degree of independence additional parameters have with amplitude and frequency.

While 200 individually recorded frequency component values is almost certainly an over-parameterizing of the perceptual space, psychophysical work suggests that merely a single sinusoid is a gross under-parameterization. \cite{bensmaia2005vibrotactile} demonstrates that high frequency vibrations are not detected simply as intensity coding; rather, at least some level of spectral complexity can be detected. How can we go about capturing the most perceptible aspects of spectral complexity with just a few parameters? Previous groups have taken a variety of approaches in the above MDS studies, such as changing rhythmicity, which could be thought of as changing spectral information over time, or adding a second frequency component. Changing waveform type can also be thought of as adding more frequency components in the form of harmonics. 

Informal experimentation suggests that white noise, even bandpassed to only higher frequencies in the fine texture regime, feels significantly different from a pure sinusoidal vibration. 
This stark difference makes ``added noise'' a potential candidate for a scalable feature of fine texture.
Recently, different types of added noise have been tested as features to continuously increase for textures that can provide directional cues \cite{bodas2019roughness}.
These methods, termed either ``injected zero drops'' or ``added white noise'' involve injecting more or less spectral noise  to an existing pattern such as a sinusoid. 
In this paper, we explore varying the breadth of spectral noise, while leaving the centroid relatively unchanged. By filtering white noise with a variable width filter, we can continuously scale from a single sinusoid to broad noise. We will refer to this characteristic as added ``irregularity''. 
We hope to understand whether this characteristic also scales continuously in perceptual space, and whether it is perceived independently of amplitude and frequency.

\subsection{Empirical Study}

Amplitude and frequency of sinusoidal textures are well established as perceptually relevant features. In this work, we present a third continuously variable texture feature, irregularity. We want to check whether it is perceptually distinct from the others, and observe how people describe its presence.

\section{Methods}

\subsection{Set Construction}

We constructed a set of variable friction textures that differed in a sinusoidal center frequency $f_0$, amplitude $A$,  and ''irregularity'' $R$. This latter quality refers to the width of the spectral content around $f_0$. Here, we directly specify spectral width via the Q-factor 
of a band pass filter applied to white noise. A larger Q-factor results in a narrower filter, and therefore less irregularity and an increasingly "pure" sine wave. 
The filter used for texture construction, shown in equation \ref{eq:filter}, uses a Q-factor that inversely depends on $R$ and a peak spectral magnitude located at $f_0$ as specified in equations \ref{eq:relationship_Q} and \ref{eq:relationship_f}. The sampling frequency $f_s$ was 100 kHz.

\begin{equation}
\displaystyle {H(z)} = \frac{\frac{\sin w_0}{2Q } -  \frac{\sin w_0}{2Q}z^{-2}}
{(1{+}\frac{\sin w_0}{2Q}) -  (2\cos w_0) z^{-1} + (1{-}\frac{\sin w_0}{2Q})z^{-2}} 
\label{eq:filter}
\end{equation}

\begin{equation}
\displaystyle { Q = \frac{1}{R} }
\label{eq:relationship_Q}
\end{equation}

\begin{equation}
\displaystyle { w_0 = \frac{2 \pi f_0}{f_s} } 
\label{eq:relationship_f}
\end{equation}

Parameters, listed in Table \ref{table_values}, each had three logarithmically scaled values, with overall ranges chosen to result in easily perceived differences.

\begin{table}[h]
\caption{Parameter Values}
\label{table_values}
\begin{center}
\begin{tabular}{ p{2cm} p{2cm}  p{1.8cm}   }
\hline
Center frequency &  amplitude &     irregularity  \\
(Hz) &              (normalized) &  (1/Q-factor) \\
\hline
\hline
150  & 0.30  & 0.067	\\
\hline
260  & 0.55  & 0.34	\\
\hline
450  & 1.0   & 1.67    \\
\end{tabular}
\end{center}
\vspace*{-10pt}
\end{table}

\begin{figure}[tbh]
      \centering     
\includegraphics[width=1\columnwidth]{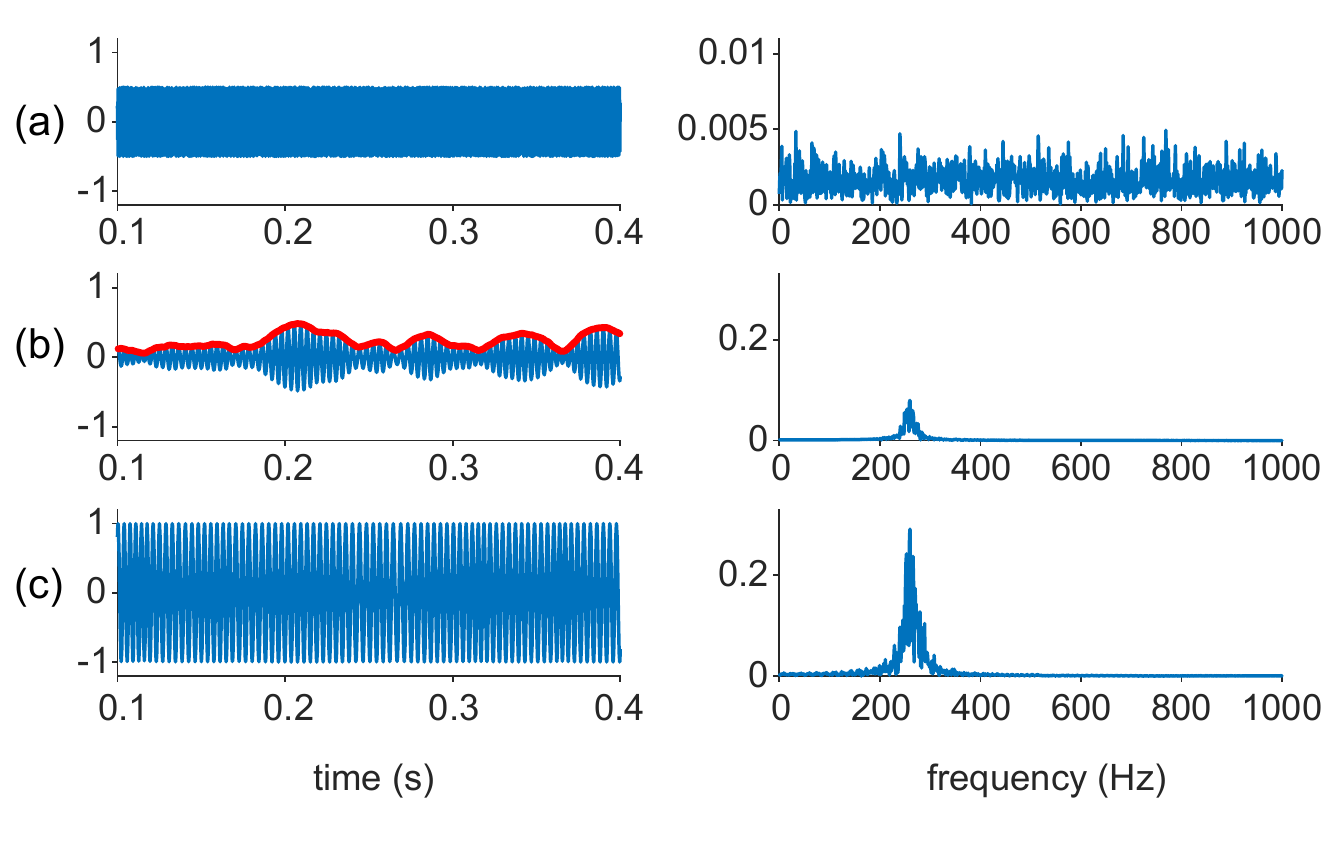}
      \caption{Demonstration of texture construction. Time domain signals are shown on in the left column for (a) broadband white noise, (b) filtered white noise and its envelope indicated in red, and (c) the filtered noise divided by its envelope. Corresponding magnitude spectra are on the right for reference. }
      \label{fig:envelope_construction}
\end{figure}

\begin{figure}[tbh]
      \centering     
\includegraphics[width=1\columnwidth]{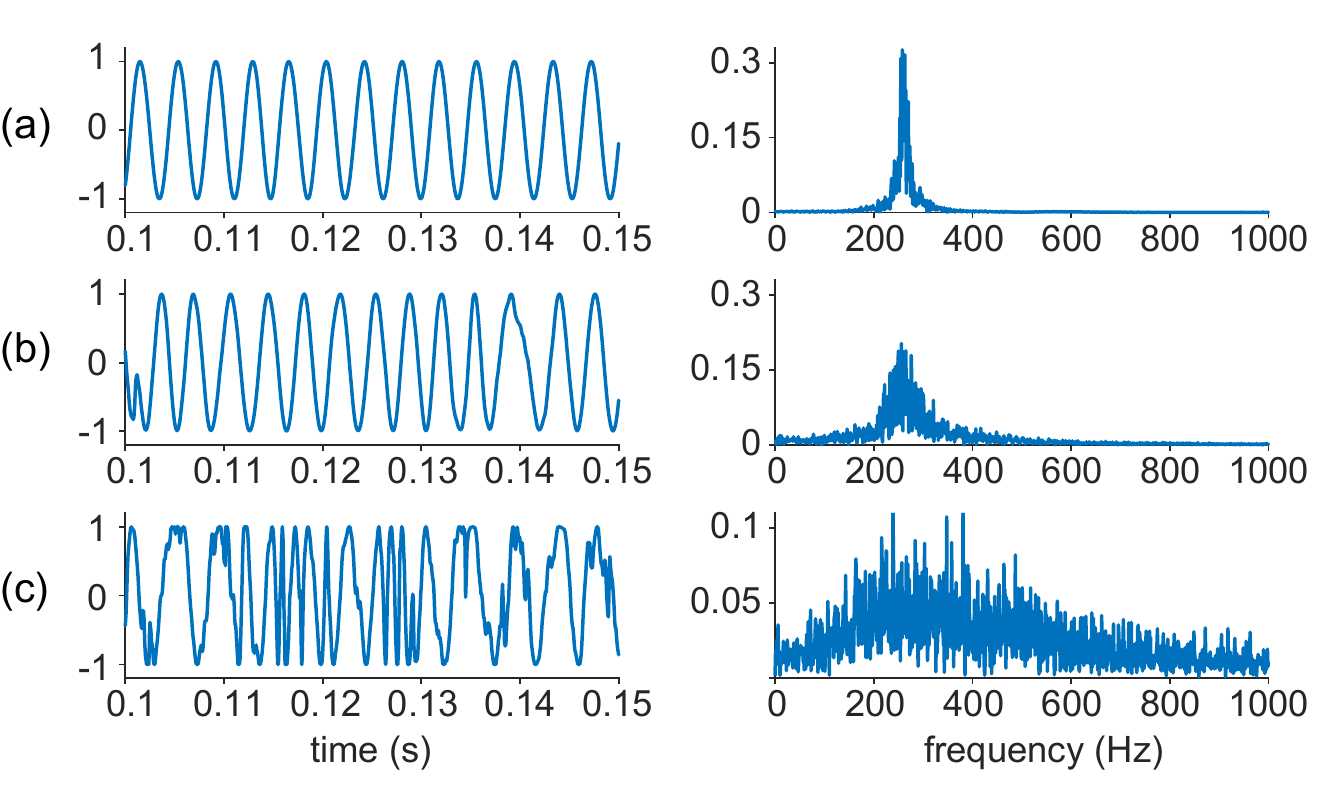}
      \caption{Time domain signals and their magnitude spectra for three different irregularity values, all with 260 Hz center frequency and amplitude of 1.}
      \label{fig:3_reg_values}
\end{figure}

Narrower band filtering initially resulted in large low frequency fluctuations in amplitude, as illustrated in Fig. \ref{fig:envelope_construction}(b). These fluctuations are problematic, as they perceptually overwhelm the finer frequency components and result primarily in the sensation of very low frequency throbbing. Additionally, they have a large impact on overall maximum amplitude, swamping any contribution of a gain term. In order to correct for these deleterious side effects, each filtered signal was divided by its envelope, calculated using the Hilbert transform via the envelope function in MATLAB 2019. The effect of this transformation on the signal in both the time domain and in frequency space is demonstrated in Fig. \ref{fig:envelope_construction}(c).  
Textures demonstrating the three different irregularity values for a 260 Hz center frequency and the maximum amplitude are shown in Fig. \ref{fig:3_reg_values}.

Dividing filtered signals by their envelope results in an maximum amplitude of one in the time domain. 
We then scale the signal down to 55\% or 30\% of its original amount for the smaller amplitude parameter values. Finally, we excluded textures with the highest frequency and lowest amplitude value from the set, as they were perceptually weak enough to be difficult to detect by the experimenters. All other amplitude and center frequency values had three variations of filter width, resulting in 24 textures total.

\subsection{Apparatus}

\begin{figure}[tbh]
      \centering     
\includegraphics[width=1\columnwidth]{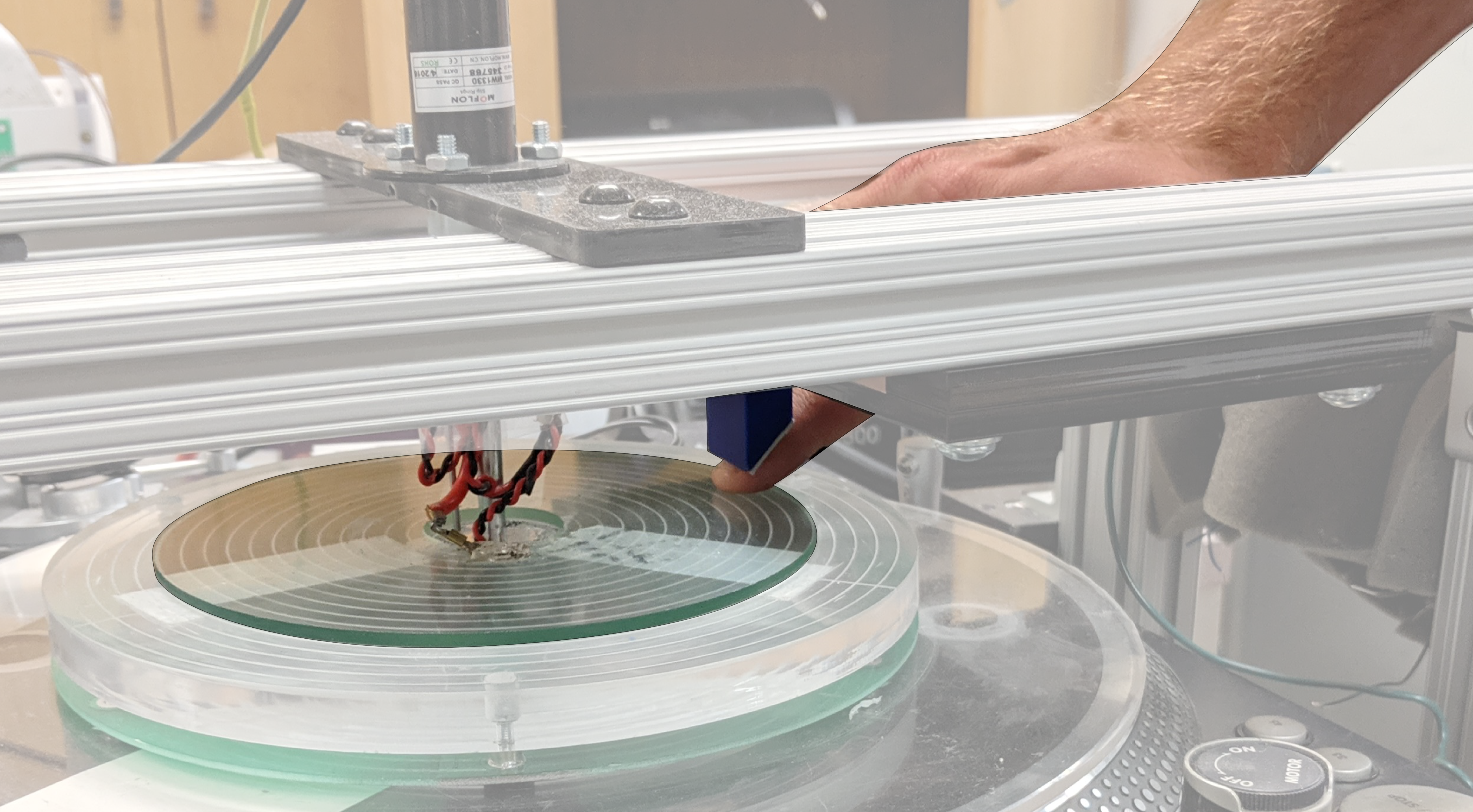}
      \caption{Image of friction-modulating display. The hand, guide block, and disk-shaped 3M screen are highlighted.}
      \label{fig:apparatus}
\end{figure}

Variable friction textures are applied via an electrostatic display using a 3M screen mounted on a turntable; detailed device specifications can be found in \cite{shultz2018application}. The turntable rotates continuously under the finger, ensuring a consistent scanning velocity across all users and negating the need to change or disrupt swipe direction. 
Users rest their right index fingernail against a guide block which ensures a consistent finger placement 
at an angle of incidence 45 degrees to the surface, approximately 60 mm from the center of the disk. The turntable rotates 33 revolutions per minute, resulting in a scanning velocity of approximately 200 mm/s perpendicular to finger orientation. See Fig. \ref{fig:apparatus} for an image of the setup. 

Textures in the above set have a unitless amplitude of up to 1 and are centered on 0. In application, friction force is controlled via the envelope of a 20 kHz commanded current applied to the display surface; therefore, textures must be offset and magnified into the appropriate current range for the given display device. 
According to \cite{shultz2018application}, the 3M screen in this apparatus has a relatively linear relationship between applied current envelope and resultant friction force in the range 1-5 mA. 
Therefore, we chose this range to correspond to the maximum normalized amplitude of 1 defined in texture construction. All textures, including those with smaller amplitudes, were centered on 3 mA to ensure consistent average friction values.

\begin{figure}[tbh]
      \centering     
\includegraphics[width=1\columnwidth]{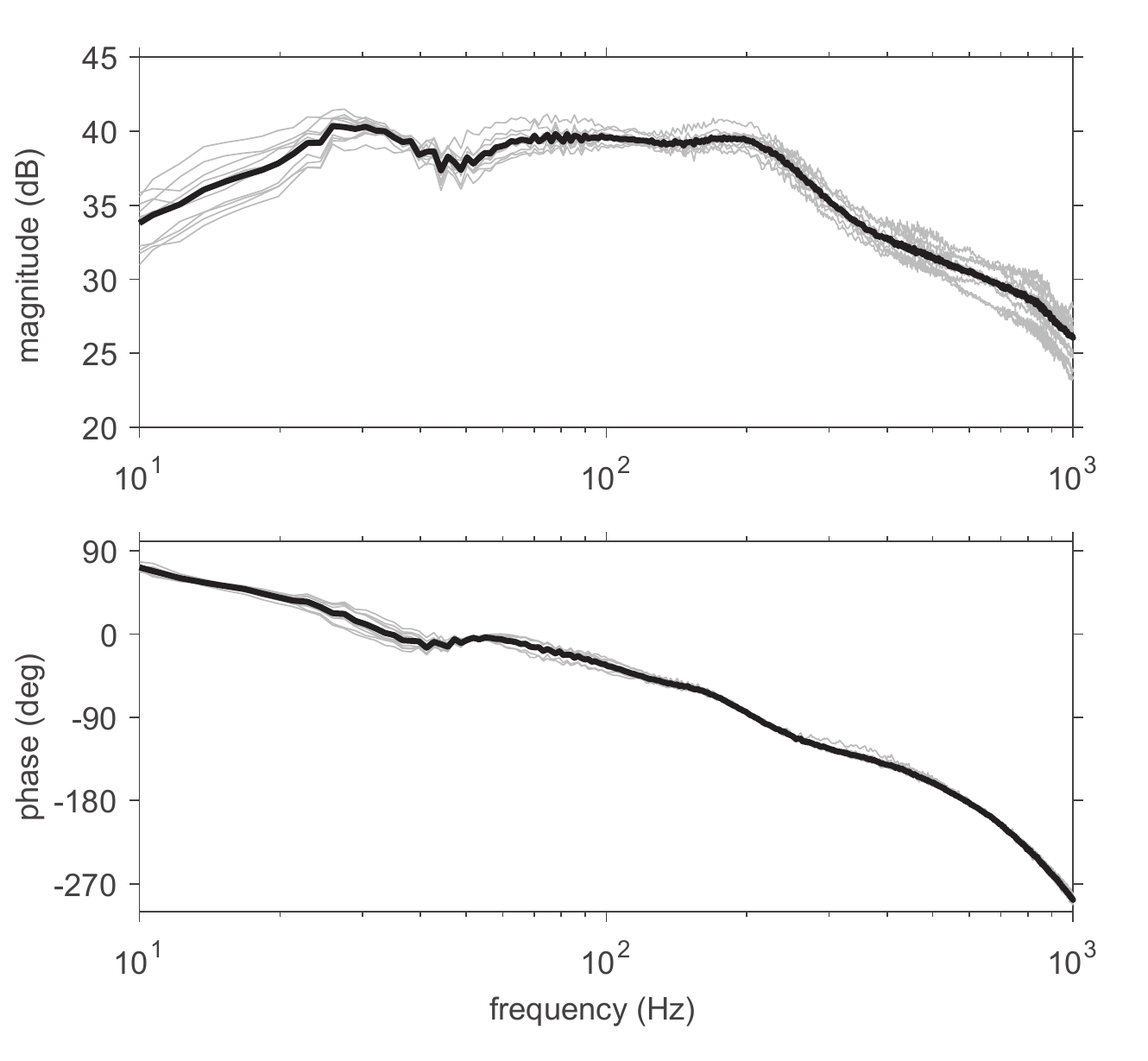}
      \caption{Bode plot for commanded friction vs LDV velocity measurements. 10 trials are shown in grayscale, while the average result is in black.}
      \label{fig:commanded_vs_measured}
\end{figure}

While we assume a linear relationship between commanded signal and applied friction force, their relationship to actual finger velocity for this setup has not been explored in depth.
In order to understand how our applied friction force translates to actual movement of the skin, we performed a frequency sweep from 10 to 1000 Hz for a constant sinusoidal current signal bounded by 1 to 5 mA. A Polytec LDV (IVS-500) measured velocity of the side of the finger a few millimeters above surface contact, and perpendicular to finger sliding direction.
Fig. \ref{fig:commanded_vs_measured} shows
data collected from the lead author's finger.
For frequencies between 25 and 250 Hz, we can see that the response is fairly flat, while velocity amplitude begins to roll off above approximately 250 Hz.

\subsection{Psychophysical Procedure}

\begin{figure}[tbh]
      \centering     
\includegraphics[width=.95\columnwidth]{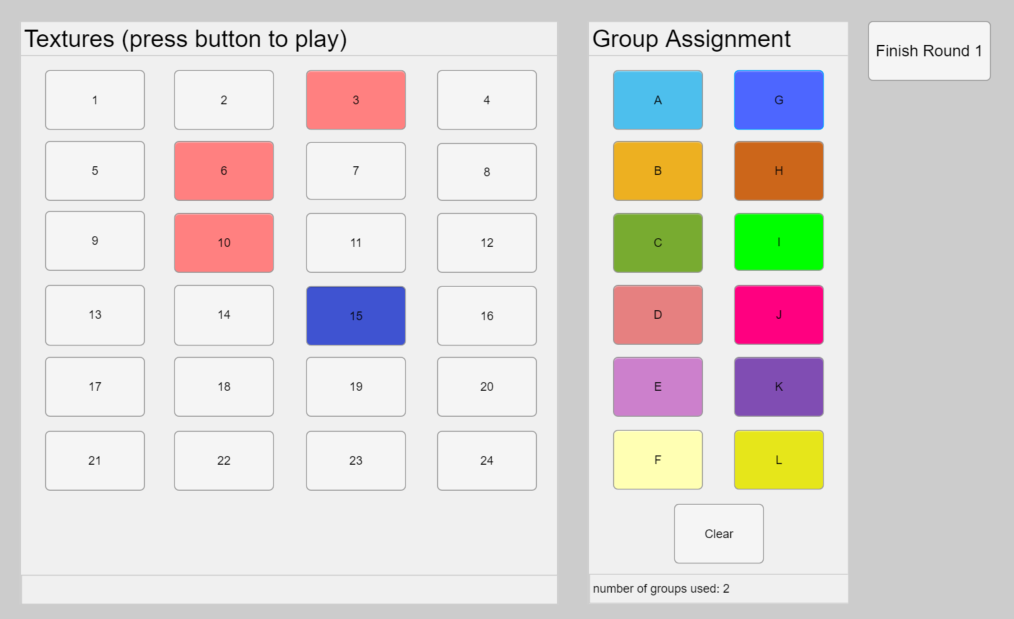}
      \caption{User interface for subject trials. In this example, textures 3, 6, and 10 are assigned to group D, texture 15 is assigned to group G, and the rest of the textures have yet to be assigned.}
      \label{fig:GUI}
\end{figure}

17 people (one left handed) ranging in age 18-35 (median age 26) participated in this study. 
This work was approved by the Northwestern Institutional Review Board, and all participants were reimbursed for their time. 
When interacting with the apparatus, participants placed their right index finger on the display surface and actuated texture playback with their left hand on a separate tablet displaying the user interface shown in Fig. \ref{fig:GUI}. They also listened to pink noise on noise cancelling headphones during texture playback in order to ensure that any audio effects coming off their finger did not influence their haptic perception. Texture order in the user interface was randomly scrambled between participants to ensure that button location did not affect grouping behavior across the population. 

At the outset of the experiment,
participants were asked to first feel the entire set of 24 textures. This helped them become comfortable with the experimental setup and ensured that they were acquainted with entire range of textures. During this time, any remaining questions regarding the user interface could be addressed, as well as any ergonomic changes such as chair height or arm support. 

Following this training period, participants began round 1 of the experiment.
They were instructed to assign all textures to groups, so that textures within a group were similar to one another, and textures in different groups more dissimilar. 
A given texture was assigned to a group by first selecting and playing that texture from the left side of the GUI, and then selecting a group assignment from the right side. 
Textures could be played as many times as desired and reassigned to different groups for the duration of the experimental round. 
Once the participant was satisfied with their group assignments, they alerted the experimenter that they were finished.

A single round of grouping is somewhat disadvantaged in that for a given person, we cannot extract any gradation of similarities between textures \cite{okamoto2012psychophysical, klatzky1993cognitive}.
Therefore, 
once all textures were assigned to an initial group, participants proceeded to round 2 where they were asked to reduce the number of groups by roughly half by combining more similar groups. In a final third round, they reduced the number of groups yet again. At the end of round 3, they gave each of their remaining groups a name or phrase to describe the common characteristic of textures within each group. After observing group names that encompassed multiple characteristics for the first few experiments, 
group names were also collected for participants \#7-17 following the second round.

\section{Results}

\subsection{Multidimensional Scaling}

\begin{figure}[tbh]
      \centering     
\includegraphics[width=1\columnwidth]{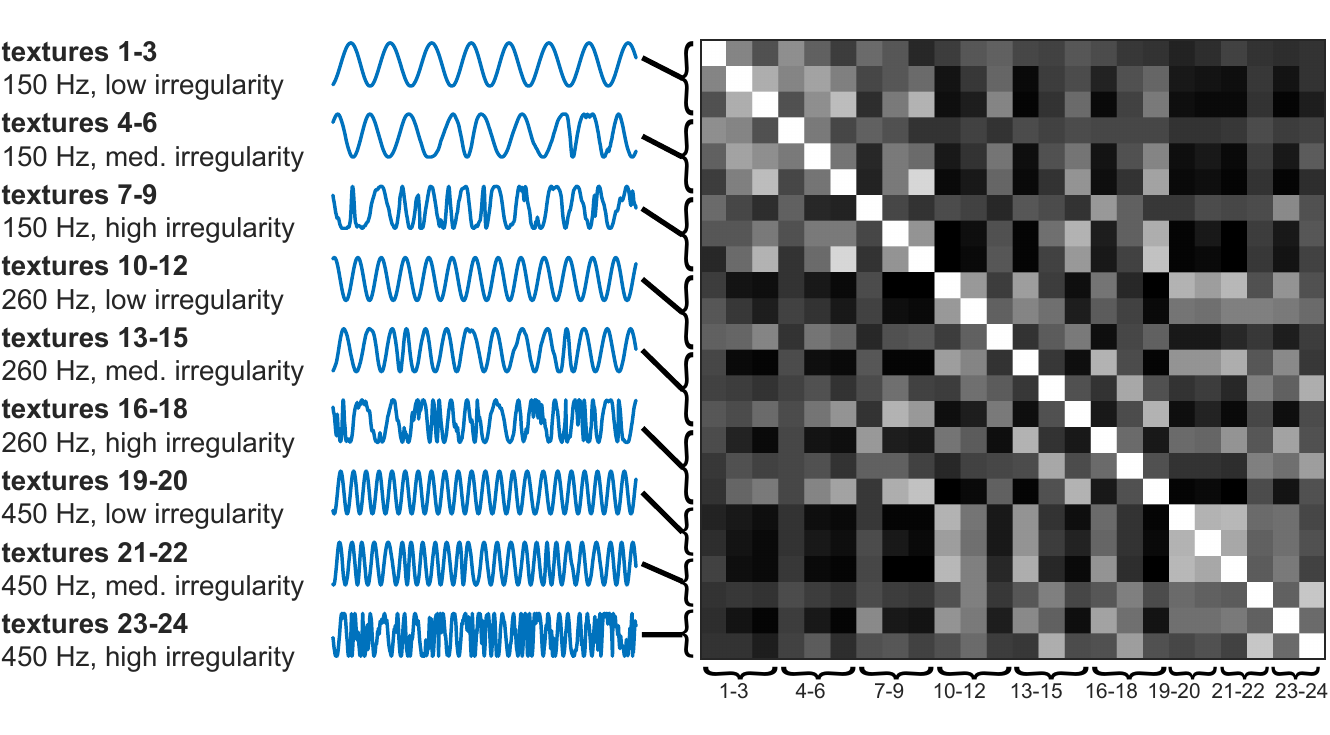}
      \caption{The similarity matrix for all 24 textures. Textures 1-18 have three different amplitude values for each example shown, increasing in sequential order, while the lowest amplitudes are omitted for the 450 Hz textures. }
      \label{fig:similarityMatrix}
\end{figure}

Three rounds of increasingly coarse grouping assignments were used to construct a matrix tabulating the perceived similarities of all textures to each other. 
Using a ranked point assignment similar to that in \cite{klatzky1993cognitive}, textures grouped together in earlier experimental rounds received more points than those grouped subsequently, resulting in zero (never grouped) to three (grouped in the first round) points from each participant for each pairing. 
Points were summed across all participants, resulting in the similarity matrix shown in Fig. \ref{fig:similarityMatrix}. The similarity relationships in this matrix were then visualized as distances using Multidimensional Scaling (MDS).

\begin{figure}[tbh]
      \centering     
\includegraphics[width=1\columnwidth]{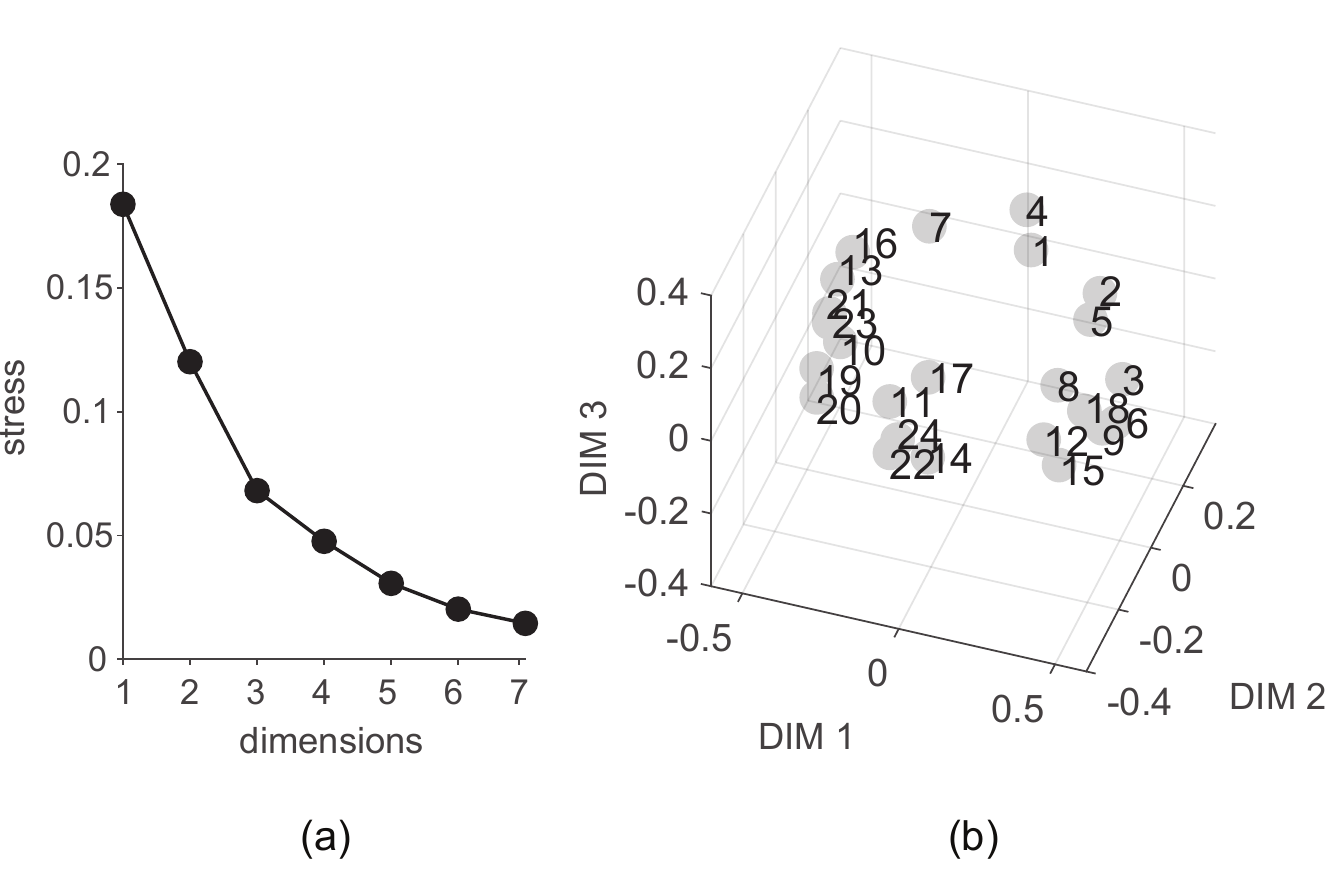}
      \caption{Results for MDS. (a) Scree plot, with elbow observed at 3 dimensions. (b) 3-dimensional perceptual space with labeled textures. }
      \label{fig:screeMDS}
\end{figure}

A nonmetric MDS analysis was performed using Krushal's stress criterion 1,
where distances in MDS space decrease monotonically with increased similarity. 
Determining the appropriate number of dimensions for this space is often nontrivial; fully satisfying the stress criterion depends not only on how many factors are actually responsible for perception, but also on any noise in subject responses. 
In other words, unless there are as many dimensions as there are stimuli, there will always be some stress (i.e. warping) of stimuli locations when fitting perceived similarities to a limited-dimension space. 
A common method for distinguishing relevant dimensions from subject-related noise is to look for a
characteristic "knee" in the stress value. This abrupt flattening of the stress for increasing numbers of dimensions indicates that further dimensions may be due primarily to noise in the data, and would give diminishing returns with regards to explaining meaningful differences in similarities. 
Fig. \ref{fig:screeMDS} (a) shows a slight knee at 3 dimensions for our data, suggesting this as an appropriate number,
although the bend is far from dramatic. 
Another commonly used metric to determine
number of dimensions is to simply choose a maximum acceptable stress cutoff, often at 0.15 \cite{nees2019perceptions}, which three dimensions more than satisfies here. 
The slightly spherical nature of the 3D solution in Fig. \ref{fig:screeMDS} (b) suggests that three dimensions does not perfectly encompass the entirety of perceived dimensions \cite{buja2002visualization}, but adding further dimensions becomes prohibitive to visualization and interpretation. 

\begin{figure*}[hbtp]
      \centering     
\includegraphics[width=1\textwidth]{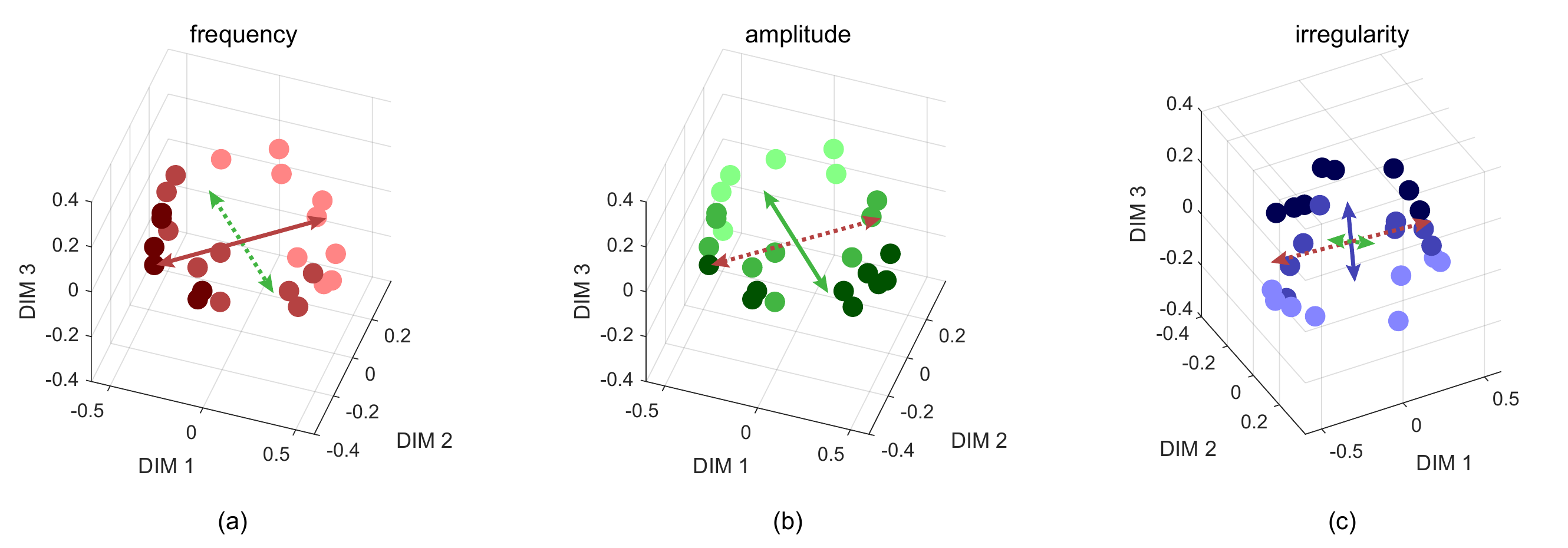}
      \caption{Projections of 3-dimensional MDS plots, color coded for each parameter indicated in the titles. Marker darkness reflects increasing values of each parameter. Plots (a) and (b) are projected perpendicular to the irregularity vector axis, while (c) is rotated to show the irregularity gradient. }
      \label{fig:mdsPlots}
\end{figure*}

We can observe how the engineering parameters, i.e. frequency, amplitude, and irregularity, map into this MDS space. Fig. \ref{fig:mdsPlots} shows the same perceptual space in all three plots, but with textures shaded differently to reflect the three values of each parameter.
Corresponding vectors for each parameter are also included. These vectors, calculated in equation  \ref{eq:gradients}, align with the direction of greatest change for a given parameter,
and their length corresponds with the influence that parameter has on perception; see \cite{tiest2006analysis} for a full derivation.
\begin{equation}
\displaystyle { {\vec{p}} = \frac{\sum{q_i\vec{x}_i}}{\sum{q_i^2}} } 
\label{eq:gradients}
\end{equation}
Here, a given vector $\vec{p}$ for a particular parameter $q$ depends 
on each $i$th texture's parameter value $q_i$ and position in MDS space $\vec{x}_i$, summed over all textures.
The first two plots in Fig. \ref{fig:mdsPlots} are oriented perpendicular to the irregularity vector and so show only the frequency and amplitude vectors, while the third plot is rotated to show the ordered distribution of irregularity values.

\subsection{Perceived Parameters}

Angles between each pair of the three engineering parameter vectors in MDS perceptual space are reported in Table \ref{table_angles}. While no pair of vectors is precisely orthogonal, as would be expected from perceptual independence, they are far from co-linear, and collectively, the three engineering vectors cover the three-dimensional MDS space. These results indicated that the engineering parameters are perceptually distinguishable but do not precisely align with the dimensions underlying texture-similarity judgments.

\begin{table}[h]
\caption{Angles between engineering parameter vectors}
\label{table_angles}
\begin{center}
\begin{tabular}{ p{3cm} p{1cm}      }
\hline
{}                     &  angle \\
\hline
\hline
frequency-amplitude    & $128^{\circ}$ 	\\
\hline
amplitude-irregularity & $74^{\circ}$ \\
\hline
irregularity-frequency & $109^{\circ}$     \\
\end{tabular}
\end{center}
\vspace*{-10pt}
\end{table}

While the distances between stimuli have clear meaning in MDS space, the rotational orientation of the entire set of stimuli does not necessarily. However, the longest section of space, corresponding with the most perceived difference, is oriented along dimension one. 
This invites speculation as to whether the MDS dimensions, particularly dimension one, are a truer representation of the ``perceived'' parameters, in contrast with the engineering parameters. 
Table \ref{table_correlations} shows the correlations for the MDS dimensions with each engineering parameter; stronger relationships are highlighted in bold. All dimensions are somewhat correlated with all three parameters, although to differing degrees.
Dimension one appears to represent an inverse relationship between frequency and amplitude, with very little correlation with irregularity, while Dimension two is primarily correlated with irregularity. Dimension three depends most strongly on irregularity and amplitude.  

\begin{table}[h]
\caption{Correlations between parameters and MDS dimensions}
\label{table_correlations}
\begin{center}
\begin{tabular}{ p{1cm} p{2cm} p{2cm}  p{2cm}   }
\hline
{} & frequency  & amplitude  &  irregularity  \\
\hline
\hline
Dim 1 & \textbf{-0.669}  & \textbf{0.480}   & -0.121	\\
\hline
Dim 2 & -0.447  & -0.420  & \textbf{-0.534}	\\
\hline
Dim 3 & -0.305  & \textbf{-0.562}  & \textbf{0.644}    \\
\end{tabular}
\end{center}
\vspace*{-10pt}
\end{table}

\subsection{Semantic Analysis}

\begin{figure}[tbh]
      \centering     
\includegraphics[width=1\columnwidth]{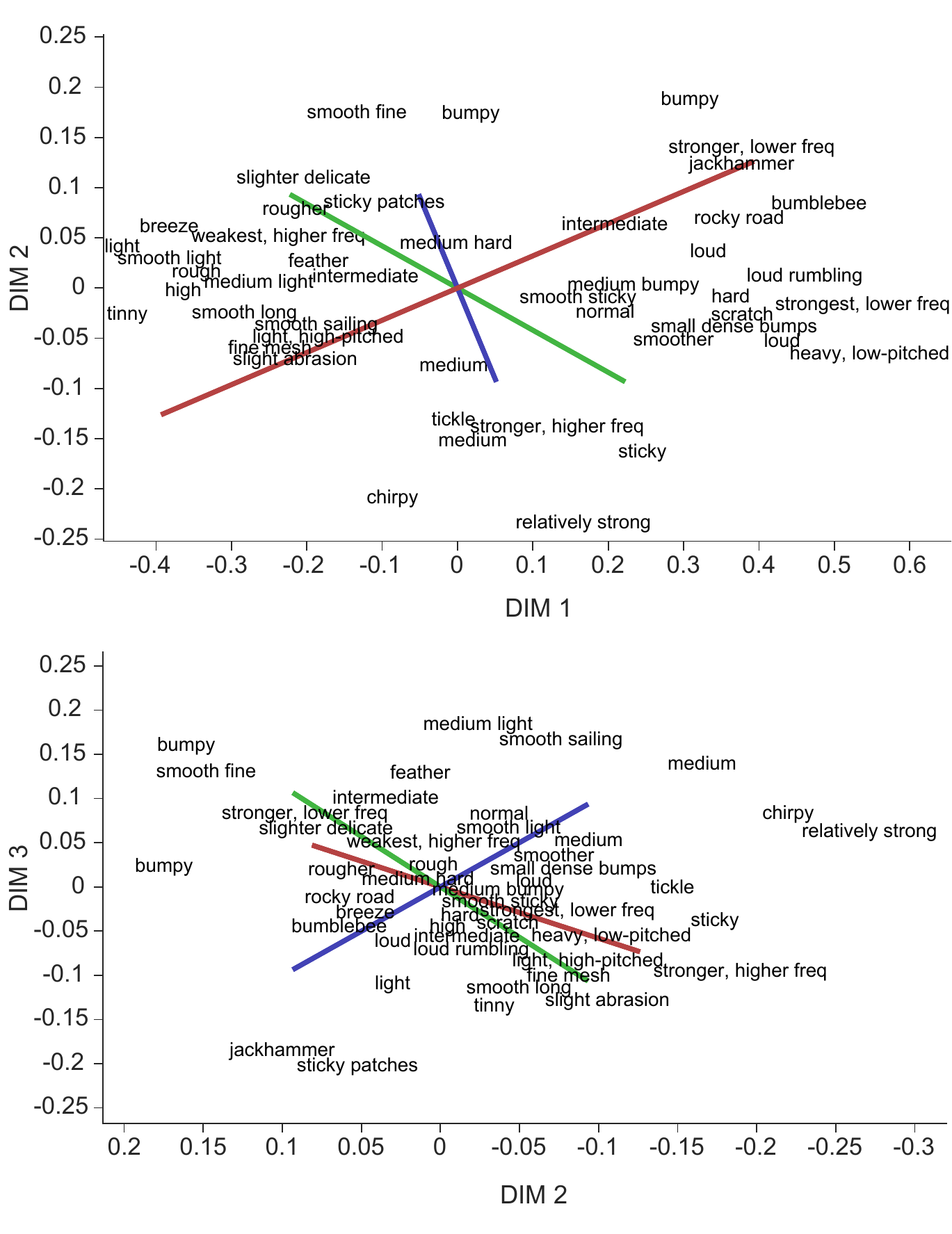}
      \caption{Projections of MDS space onto dimensions 1 and 2, and dimensions 2 and 3. Engineering parameter vectors are shown in red, green and blue, with the same color assignments from Fig. \ref{fig:mdsPlots}. Group names are shown at the mean location between all textures in a group. Some words have been shifted slightly for legibility. }
      \label{fig:words}
\end{figure}

In addition to observing how the engineering parameters are distributed throughout the perceptual space, 
we analyzed qualitative descriptions of 
the space using the assigned texture group names.
Participants were initially asked to give descriptive names to texture groups only on the final round of grouping, either as a word or short phrase. After the first 6 subjects, it was apparent that the final grouping stage was too late; people were naming groups with longer phrases that encompassed several types, such as ``rough/scratchy/low/deep'' or ``strong or bass or medium freq + high amplitude''. Therefore, we asked the subsequent 11 participants to name their groups following
the second round of grouping. This resulted in names that
generally encompassed one to two traits, such as ``chirpy'' or ``stronger, higher frequency'', respectively. All group names from the second round of grouping are distributed in Fig. \ref{fig:words}. Each name is shown at the average location of all the textures that were members of that group.

Many of the words along dimension one are directly related to the engineering parameters amplitude and frequency, such as ``strongest, lower frequency'' and ``weakest, higher frequency''. Other words, such as ``jackhammer'' and ``slighter delicate'', suggest a variety of perceived intensities. Dimensions two and three have fewer group names spread along their axes, and the perceptions underlying these words, such as``chirpy" and ``breeze'', are less immediately obvious.

\section{Discussion}

The three engineering parameters, frequency, amplitude, and irregularity, are all well represented in MDS space, as indicated by the corresponding vectors calculated  in equation 3.  In addition, the angles between vector pairs indicate substantial independence, if not orthogonality. 
These results suggest that the engineering parameters could be good candidates for adjustable controls for texture design.  
The relative length of these vectors, as computed from the distribution of the textures along them, indicates their importance to perceived textural similarity, which could be used to scale the controls to maximize salient contributions.

While clearly differentiated in the MDS solution, the parameter vectors are not aligned with the three emergent perceptual dimensions. This is not necessarily an indication that the actual perceived features are different from the engineered ones, as MDS maps out inter-stimulus distances and not orientation.
However, the most differentiated stretch of the MDS space, oriented along dimension one, appears to be a perceptual conflation of frequency and amplitude. Table \ref{table_correlations} shows that this dimension has an inverse dependence on frequency and amplitude, and almost no correlation with irregularity. In Fig. \ref{fig:words}, we can see that subjects' group names also refer to combinations of these parameters with terms along dimension one such as ``low-pitched" and ``high-pitched". Pitch, i.e. perceived tactile frequency, specifically has been shown to depend on an inverse relationship between frequency and amplitude \cite{von1959synchronism}, and can be identified even when there are multiple frequency components present \cite{friesen2018single,hwang2017perceptual}.

While subjects' group names were quite directly related to frequency, amplitude, and combinations of the two, discernible patterns in names along the irregularity axis were not as clear. Creative names such as “bumblebee” and “chirpy” hint at substantive experiences, but are not easily interpreted by other readers.
Subjects do not appear to innately have the vocabulary to describe this type of parameter change. A better approach may be to use this as inspiration to provide them with a predefined vocabulary list from which they can select or rank terms.

The primary goal of this work was to find distinguishable features of texture that could be independently adjusted and continuously scaled. Future work is needed to determine whether and how well people can use the engineering parameters in this study to adjust textural features.
Examples of such adjustments include
intentionally enhancing a chosen trait, or trying to perceptually match another texture using a combination of all parameter settings. 
Would users be able to navigate a 3-dimensional design space faster or more accurately with the original engineering parameters, or would the emergent MDS dimensions provide more intuitive control? 
Adjustment ''knobs" representing emergent dimensions might consist of linear combinations of the engineering parameters, or other functions of the parameters that represent established characteristics such as tactile pitch.
An additional unknown is how coarse the step
size should be for a given parameter implemented in adjustable controls.

Characteristics such as temperature, deformability, and pile all contribute to the rich breadth of fine textures we can distinguish in the real world. However, even within the limitations of a screen that can only modulate friction forces, it is conceivable that there are more independent textural characteristics not addressed by our three engineering parameters. For this reason, future experiments should examine whether there are friction modulated fine textures that cannot be perceptually matched by our texture generation scheme. This would take us one step further toward understanding the full gamut of friction modulated textures, and how our 3-parameter space fits within it. 

\section{Conclusion}

We constructed a 3-dimensional fine texture design space for variable friction haptic surface displays. Engineering design parameters included frequency and amplitude of sinusoidal changes in friction, as well as an irregularity term proportional to spectral width. Via multidimensional scaling, these parameters show up in perceptual space relatively orthogonal to each other. Perceptual dimensions appear to be combinations of the engineering parameters, with the primary dimension an inverse relationship between frequency and amplitude, the second dimension most dependent on irregularity, and the third a function of all three parameters. Analysis of texture group names given by subjects suggest that people have no problem identifying pitch-like and volume-like features, but struggle to cohesively name injections of irregularity.

\appendices


\section*{Acknowledgment}

This work was generously made possible by funding from the NSF Grant IIS-1302422.

\ifCLASSOPTIONcaptionsoff
  \newpage
\fi


\bibliographystyle{./IEEEtran} 
\bibliography{texture.bib}

\begin{thebibliography}{10}
\providecommand{\url}[1]{#1}
\csname url@samestyle\endcsname
\providecommand{\newblock}{\relax}
\providecommand{\bibinfo}[2]{#2}
\providecommand{\BIBentrySTDinterwordspacing}{\spaceskip=0pt\relax}
\providecommand{\BIBentryALTinterwordstretchfactor}{4}
\providecommand{\BIBentryALTinterwordspacing}{\spaceskip=\fontdimen2\font plus
\BIBentryALTinterwordstretchfactor\fontdimen3\font minus
  \fontdimen4\font\relax}
\providecommand{\BIBforeignlanguage}[2]{{%
\expandafter\ifx\csname l@#1\endcsname\relax
\typeout{** WARNING: IEEEtran.bst: No hyphenation pattern has been}%
\typeout{** loaded for the language `#1'. Using the pattern for}%
\typeout{** the default language instead.}%
\else
\language=\csname l@#1\endcsname
\fi
#2}}
\providecommand{\BIBdecl}{\relax}
\BIBdecl

\bibitem{watanabe1995method}
T.~Watanabe and S.~Fukui, ``A method for controlling tactile sensation of
  surface roughness using ultrasonic vibration,'' in \emph{Robotics and
  Automation, 1995. Proceedings., 1995 IEEE International Conference on},
  vol.~1.\hskip 1em plus 0.5em minus 0.4em\relax IEEE, 1995, pp. 1134--1139.

\bibitem{winfield2007t}
L.~Winfield, J.~Glassmire, J.~E. Colgate, and M.~Peshkin, ``T-pad: Tactile
  pattern display through variable friction reduction,'' in \emph{EuroHaptics
  Conference, 2007 and Symposium on Haptic Interfaces for Virtual Environment
  and Teleoperator Systems. World Haptics 2007. Second Joint}.\hskip 1em plus
  0.5em minus 0.4em\relax IEEE, 2007, pp. 421--426.

\bibitem{biet2007squeeze}
M.~Biet, F.~Giraud, and B.~Lemaire-Semail, ``Squeeze film effect for the design
  of an ultrasonic tactile plate,'' \emph{Ultrasonics, Ferroelectrics and
  Frequency Control, IEEE Transactions on}, vol.~54, no.~12, pp. 2678--2688,
  2007.

\bibitem{shultz2018application}
C.~Shultz, M.~Peshkin, and J.~E. Colgate, ``The application of tactile,
  audible, and ultrasonic forces to human fingertips using broadband
  electroadhesion,'' \emph{IEEE transactions on haptics}, vol.~11, no.~2, pp.
  279--290, 2018.

\bibitem{vardar2017effect}
Y.~Vardar, B.~G{\"u}{\c{c}}l{\"u}, and C.~Basdogan, ``Effect of waveform on
  tactile perception by electrovibration displayed on touch screens,''
  \emph{IEEE transactions on haptics}, vol.~10, no.~4, pp. 488--499, 2017.

\bibitem{katz1925world}
D.~Katz, ``The world of touch. translated by le krueger,'' 1925.

\bibitem{lederman2009haptic}
S.~J. Lederman and R.~L. Klatzky, ``Haptic perception: A tutorial,''
  \emph{Attention, Perception, \& Psychophysics}, vol.~71, no.~7, pp.
  1439--1459, 2009.

\bibitem{manfredi2014natural}
L.~R. Manfredi, H.~P. Saal, K.~J. Brown, M.~C. Zielinski, J.~F. Dammann, V.~S.
  Polashock, and S.~J. Bensmaia, ``Natural scenes in tactile texture,''
  \emph{Journal of neurophysiology}, vol. 111, no.~9, pp. 1792--1802, 2014.

\bibitem{meyer2015modeling}
D.~J. Meyer, M.~A. Peshkin, and J.~E. Colgate, ``Modeling and synthesis of
  tactile texture with spatial spectrograms for display on variable friction
  surfaces,'' in \emph{World Haptics Conference (WHC), 2015 IEEE}.\hskip 1em
  plus 0.5em minus 0.4em\relax IEEE, 2015, pp. 125--130.

\bibitem{cholewiak2009frequency}
S.~A. Cholewiak, K.~Kim, H.~Z. Tan, and B.~D. Adelstein, ``A frequency-domain
  analysis of haptic gratings,'' \emph{IEEE Transactions on Haptics}, vol.~3,
  no.~1, pp. 3--14, 2009.

\bibitem{meyer2016tactile}
D.~J. Meyer, M.~A. Peshkin, and J.~E. Colgate, ``Tactile paintbrush: A
  procedural method for generating spatial haptic texture,'' in \emph{Haptics
  Symposium (HAPTICS), 2016 IEEE}.\hskip 1em plus 0.5em minus 0.4em\relax IEEE,
  2016, pp. 259--264.

\bibitem{friesen2018single}
R.~F. Friesen, R.~L. Klatzky, M.~A. Peshkin, and J.~E. Colgate, ``Single pitch
  perception of multi-frequency textures,'' in \emph{2018 IEEE Haptics
  Symposium (HAPTICS)}.\hskip 1em plus 0.5em minus 0.4em\relax IEEE, 2018, pp.
  290--295.

\bibitem{hwang2017perceptual}
I.~Hwang, J.~Seo, and S.~Choi, ``Perceptual space of superimposed
  dual-frequency vibrations in the hands,'' \emph{PloS one}, vol.~12, no.~1, p.
  e0169570, 2017.

\bibitem{holliins1993perceptual}
M.~Holliins, R.~Faldowski, S.~Rao, and F.~Young, ``Perceptual dimensions of
  tactile surface texture: A multidimensional scaling analysis,''
  \emph{Perception \& psychophysics}, vol.~54, no.~6, pp. 697--705, 1993.

\bibitem{okamoto2012psychophysical}
S.~Okamoto, H.~Nagano, and Y.~Yamada, ``Psychophysical dimensions of tactile
  perception of textures,'' \emph{IEEE Transactions on Haptics}, vol.~6, no.~1,
  pp. 81--93, 2012.

\bibitem{tiest2006analysis}
W.~M.~B. Tiest and A.~M. Kappers, ``Analysis of haptic perception of materials
  by multidimensional scaling and physical measurements of roughness and
  compressibility,'' \emph{Acta psychologica}, vol. 121, no.~1, pp. 1--20,
  2006.

\bibitem{vardar2019fingertip}
Y.~Vardar, C.~Wallraven, and K.~J. Kuchenbecker, ``Fingertip interaction
  metrics correlate with visual and haptic perception of real surfaces,'' in
  \emph{2019 IEEE World Haptics Conference (WHC)}.\hskip 1em plus 0.5em minus
  0.4em\relax IEEE, 2019, pp. 395--400.

\bibitem{mun2019perceptual}
S.~Mun, H.~Lee, and S.~Choi, ``Perceptual space of regular homogeneous haptic
  textures rendered using electrovibration,'' in \emph{2019 IEEE World Haptics
  Conference (WHC)}.\hskip 1em plus 0.5em minus 0.4em\relax IEEE, 2019, pp.
  7--12.

\bibitem{ternes2008designing}
D.~Ternes and K.~E. Maclean, ``Designing large sets of haptic icons with
  rhythm,'' in \emph{International Conference on Human Haptic Sensing and Touch
  Enabled Computer Applications}.\hskip 1em plus 0.5em minus 0.4em\relax
  Springer, 2008, pp. 199--208.

\bibitem{bernard2018harmonious}
C.~Bernard, J.~Monnoyer, and M.~Wiertlewski, ``Harmonious textures: The
  perceptual dimensions of synthetic sinusoidal gratings,'' in
  \emph{International Conference on Human Haptic Sensing and Touch Enabled
  Computer Applications}.\hskip 1em plus 0.5em minus 0.4em\relax Springer,
  2018, pp. 685--695.

\bibitem{dariosecq2020investigating}
M.~Dariosecq, P.~Pl{\'e}nacoste, F.~Berthaut, A.~Kaci, and F.~Giraud,
  ``Investigating the semantic perceptual space of synthetic textures on an
  ultrasonic based haptic tablet,'' in \emph{HUCAPP 2020}, 2020.

\bibitem{von1959synchronism}
G.~Von~B{\'e}k{\'e}sy, ``Synchronism of neural discharges and their
  demultiplication in pitch perception on the skin and in hearing,'' \emph{The
  Journal of the Acoustical Society of America}, vol.~31, no.~3, pp. 338--349,
  1959.

\bibitem{bensmaia2005vibrotactile}
S.~Bensma{\"\i}a, M.~Hollins, and J.~Yau, ``Vibrotactile intensity and
  frequency information in the pacinian system: A psychophysical model,''
  \emph{Attention, Perception, \& Psychophysics}, vol.~67, no.~5, pp. 828--841,
  2005.

\bibitem{bodas2019roughness}
P.~Bodas, R.~F. Friesen, A.~Nayak, H.~Z. Tan, and R.~Klatzky, ``Roughness
  rendering by sinusoidal friction modulation: Perceived intensity and gradient
  discrimination,'' in \emph{2019 IEEE World Haptics Conference (WHC)}.\hskip
  1em plus 0.5em minus 0.4em\relax IEEE, 2019, pp. 443--448.

\bibitem{klatzky1993cognitive}
R.~L. Klatzky, J.~Pellegrino, B.~P. McCloskey, and S.~J. Lederman, ``Cognitive
  representations of functional interactions with objects,'' \emph{Memory \&
  Cognition}, vol.~21, no.~3, pp. 294--303, 1993.

\bibitem{nees2019perceptions}
M.~Nees, ``Perceptions of terms used to describe automation in vehicles: A
  multidimensional scaling study,'' 2019.

\bibitem{buja2002visualization}
A.~Buja and D.~F. Swayne, ``Visualization methodology for multidimensional
  scaling,'' \emph{Journal of Classification}, vol.~19, no.~1, p.~7, 2002.

\end{thebibliography}

%


\begin{IEEEbiography}[{\includegraphics[width=1in,height=1.25in,clip,keepaspectratio]{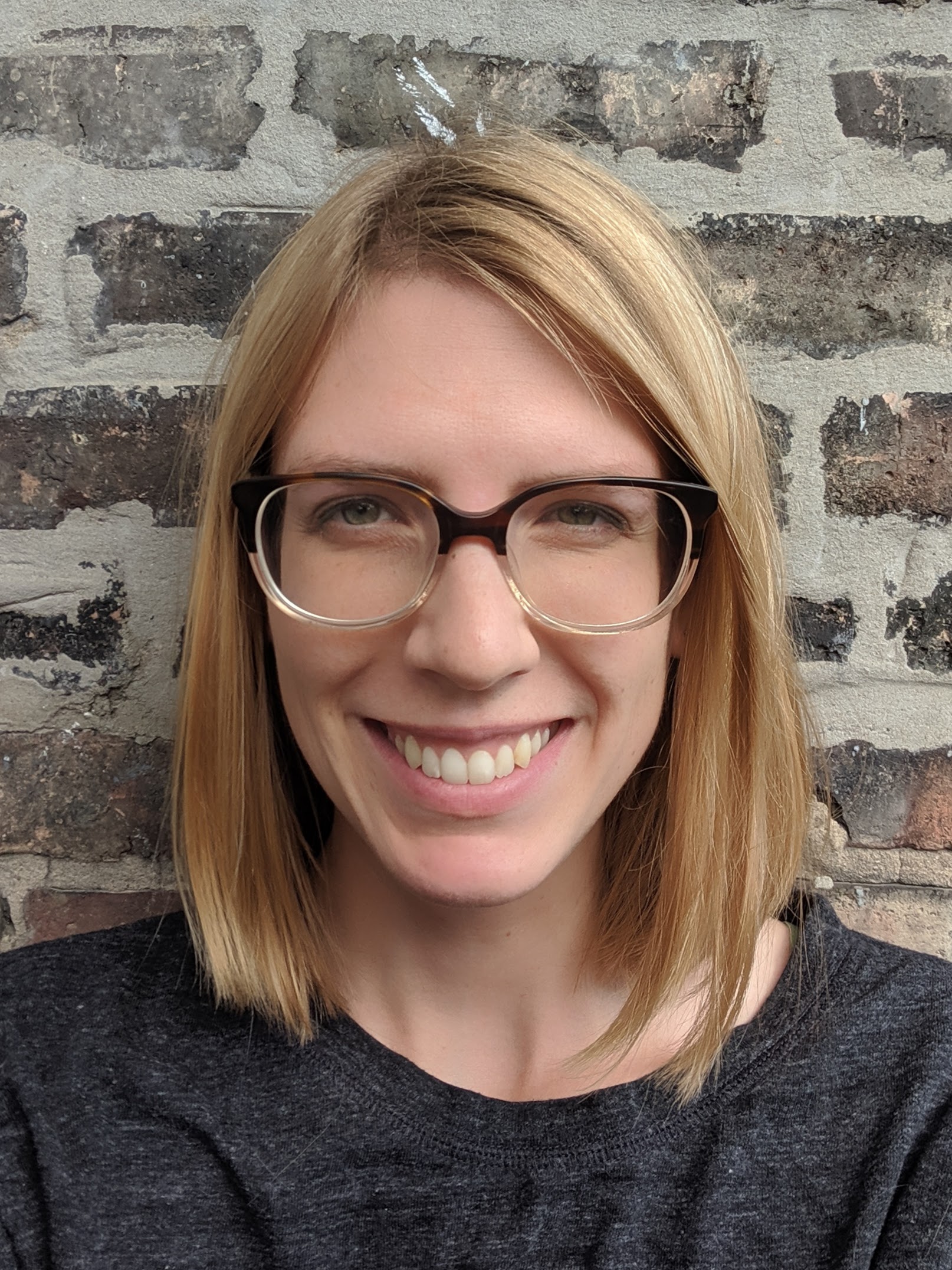}}]{Rebecca Fenton Friesen}
is a PhD candidate in the Mechanical Engineering Department at Northwestern University. Her work is centered on surface haptics, including work on the actuation of friction modulating screens and human perception of friction modulated texture.
\end{IEEEbiography}

\begin{IEEEbiography}[{\includegraphics[width=1in,height=1.25in,clip,keepaspectratio]{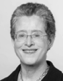}}]{Roberta L. Klatzky}
received the Ph.D. degree in cognitive psychology from Stanford University. She is the Charles J. Queenan, Jr. University Professor at Carnegie Mellon University, Pittsburgh, PA, USA, where she is a member of the Dept. of Psychology, Human–Computer Interaction Institute, and Neuroscience Institute. Her research interests include human perception and cognition, with special emphasis on spatial cognition and haptic perception. She is a Fellow of the IEEE.
\end{IEEEbiography}

\begin{IEEEbiography}[{\includegraphics[width=1in,height=1.25in,clip,keepaspectratio]{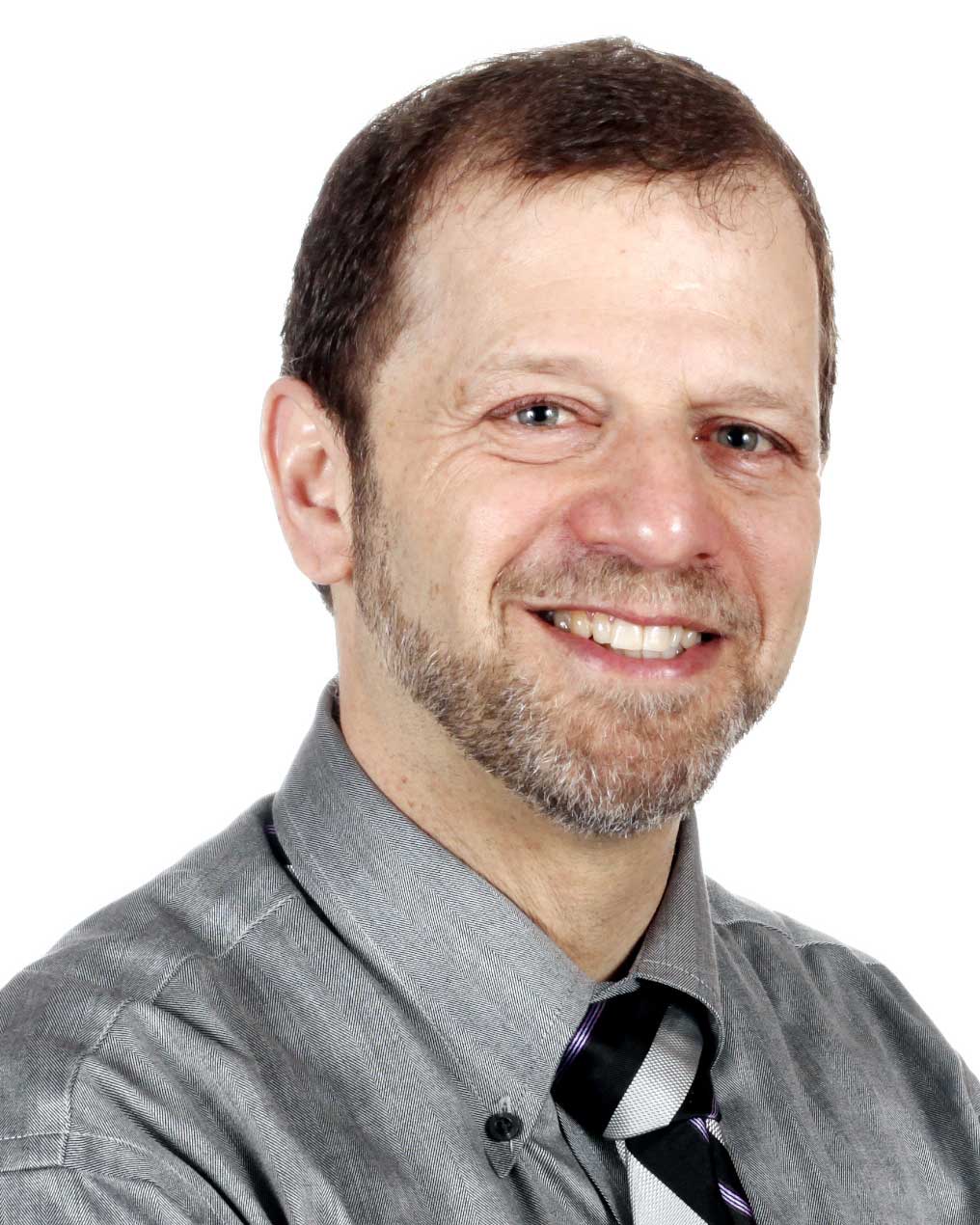}}]{Michael A. Peshkin}
 (SM’09) is a professor in the Department of Mechanical Engineering at Northwestern University (Evanston, IL.)  His research is in haptics, robotics, human–machine interaction, and rehabilitation robotics. He has cofounded four start-up companies: Mako Surgical, Cobotics, HDT Robotics, and Tanvas. He is a Fellow of the National Academy of Inventors, and (with J. E. Colgate) an inventor of cobots.
\end{IEEEbiography}


\begin{IEEEbiography}[{\includegraphics[width=1in,height=1.25in,clip,keepaspectratio]{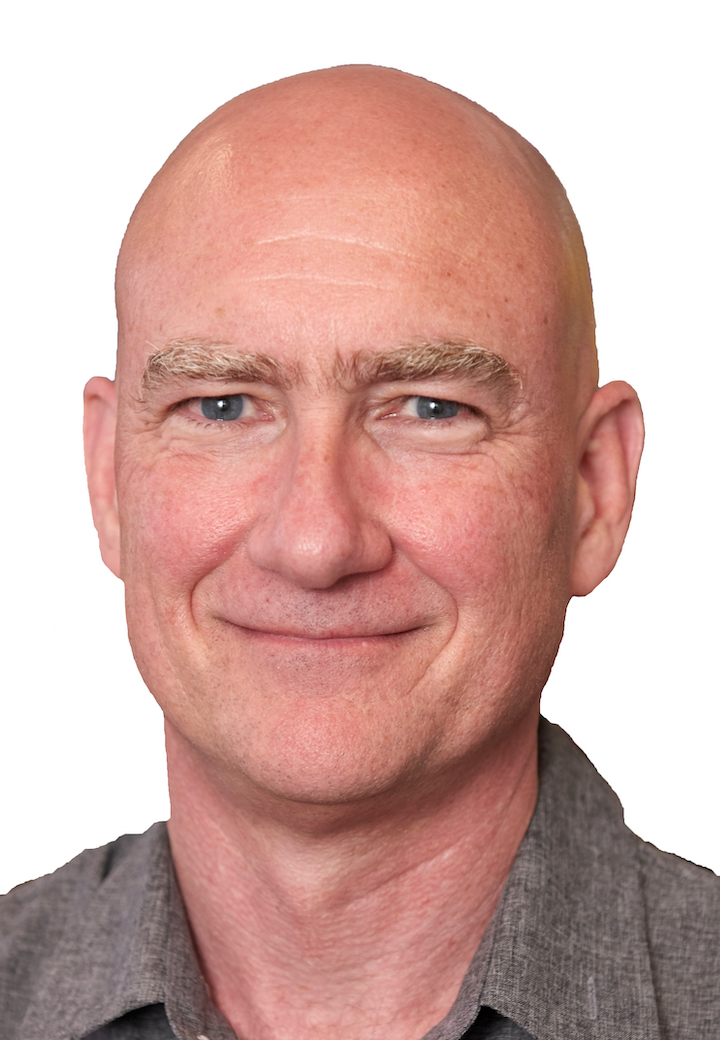}}]{J. Edward Colgate}
is the Breed University Design Professor and a member of the Department of Mechanical Engineering at Northwestern University where his laboratory focuses on haptic interface. He served as the founding Editor-in-Chief of the IEEE Transactions on Haptics and is a Fellow of the IEEE and the National Academy of Inventors.  Dr. Colgate has founded three start-up companies including Tanvas Inc., which produces advanced haptic touchscreens.
\end{IEEEbiography}




\end{document}